\documentclass[useAMS,usenatbib]{mn2e}
\usepackage{longtable}
\usepackage{amsmath,multirow,dcolumn,fancyhdr,charter,graphicx}
\usepackage{graphics}
\usepackage{color,ulem,epstopdf}

\newcommand{\kep}{{\it Kepler}}

\usepackage{tikz}
\def\checkmark{\tikz\fill[scale=0.4](0,.35) -- (.25,0) -- (1,.7) -- (.25,.15) -- cycle;} 

\title[Tides Alone Cannot Explain Kepler Planets Close to 2:1 MMR]{Tides Alone Cannot Explain Kepler Planets Close to 2:1 MMR}
\author[Silburt \& Rein]{Ari Silburt$^{1}$ \&  Hanno Rein$^{1}$ \\
$^1$Department of Environmental and Physical Sciences, University of
  Toronto Scarborough, 1265 Military Trail, Toronto, ON, M1C 1A4, Canada} 

\date{Released \today}

\begin{document}
\maketitle

\begin{abstract} 

A number of \kep{} planet pairs lie just wide of first-order mean motion resonances (MMRs).
Tides have been frequently proposed to explain these pileups, but it is still an ongoing discussion.  
We contribute to this discussion by calculating an optimistic theoretical estimate on the minimum initial eccentricity required by \kep{} planets to explain the current observed spacing, and compliment these calculations with N-body simulations.
In particular, we investigate 27 \kep{} systems having planets within $6\%$ of the 2:1 MMR, and find that the initial eccentricities required to explain the observed spacings are unreasonable from simple dynamical arguments.
Furthermore, our numerical simulations reveal \textit{resonant tugging}, an effect which conspires against the migration of resonant planets away from the 2:1 MMR, requiring even higher initial eccentricities in order to explain the current \kep{} distribution. 
Overall, we find that tides alone cannot explain planets close to 2:1 MMR, and additional mechanisms are required to explain these systems. 

\end{abstract}

%%%%%%%%%%%%%%%%%%%%%%%%%%%%%%%%%%%%%%%%%%%%%%%%%%%
%%%%%%%%%%%%%%%%%%%%%%%%%%%%%%%%%%%%%%%%%%%%%%%%%%%
%%%%%%%%%%%%%%%%%%%%%%%%%%%%%%%%%%%%%%%%%%%%%%%%%%%
\section{Introduction}
\label{sec:introduction}
The NASA \kep{} mission has been immensely successful for detecting planets outside of our solar system.
To date, it has discovered over 4,500 exoplanet candidates along with 466 multi-planet systems \citep{Akeson2013,Rowe2014}. 
A number of these systems are just wide of a mean motion resonance (MMR), which occurs when the period of one planet is an integer ratio of another. 
In particular, statistical excesses in the period distribution of \kep{} planets have been detected just wide of the 3:2 and 2:1 MMR \citep[][]{Lissauer2011,Fabrycky2014,Steffen2015}.
It is believed that in the past these planetary systems migrated into resonance via convergent migration \citep{Lee2002}, and a number of dissipative mechanisms have been proposed to slowly bring these planets out of MMR.
The most popular dissipative mechanisms to explain the observed near-resonant systems are tidal \citep{LithwickWu2012, Batygin2013, Delisle2014},  protoplanetary \citep{Rein2012b, Baruteau2013, Goldreich2014}, and planetesimal \citep{Moore2013, Chatterjee2015}. 

In this work we focus exclusively on tidal dissipation, for which there is no clear consensus on whether this mechanism alone can successfully explain the excess of near-resonant pairs near first order MMRs. 
Several authors \citep{LithwickWu2012,Batygin2013} have argued for \citep{Delisle2014} tidal dissipation, whereas others \citep{Lee2013} have argued against it. 
In particular, \citet{LithwickWu2012} first introduced the mechanism of \textit{resonant repulsion}, and showed that in the limit of low eccentricities for near-resonant planets, the space between planets grows as~$t^{1/3}$.
\citet{Batygin2013} confirmed this result.
\citet{Lee2013} then used this relationship to show that most near-resonant planet pairs cannot be explained via this mechanism, the few exceptions being small rocky planets for which tidal dissipation is particularly effective.
\citet{Delisle2014} then suggested a high eccentricity mechanism by which planets may still be able to evolve to their current positions via tides alone.

\begin{figure}
\centerline{\includegraphics[scale=0.49]{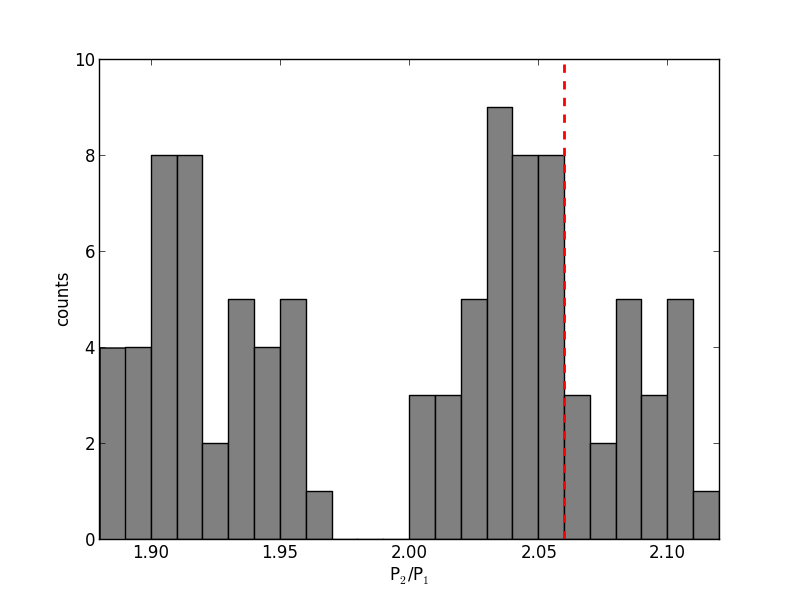}}
\caption{ \kep{} systems close to 2:1 MMR. A statistical excess is present just wide of the 2:1 MMR, and appears to decline beyond $6\%$ of the resonance, as marked by a red dotted line. 
 }
\label{fig:MMR}
\end{figure}

We contribute to this debate by developing optimistic theoretical estimates for the evolution of planets away from resonance, and compare these estimates to N-body simulations. 
We then make a statement about the likelihood of tides to explain near-resonant pairs. 
We focus on \kep{} systems within $6\%$ of the 2:1 MMR, which appears to be the natural cutoff for this statistical excess, as shown by a dotted line in Figure~\ref{fig:MMR}. 
Using this sample we will calculate the minimum required initial eccentricity to explain their current positions, given that they started in MMR and evolved under the influence of tides alone. 
In addition, we present numerical findings of {\it resonant tugging}, an effect which prevents the evolution of planets away from MMR when eccentricity is high, making it significantly more difficult to achieve the observed spacings seen today. 
Resonant tugging does not appear to have been extensively studied/accounted for in this field due to the fact that most analysis of planets in resonance have worked in the $e \ll 1$ limit, and we find that resonant tugging exclusively affects planets in MMR with moderate to high~$e$. 

Our paper is organized as follows -- in Section~\ref{sec:meth} we outline our theoretical and numerical framework, in Section~\ref{sec:results} we present the main findings of this paper, in Section~\ref{sec:Discussion} we present a discussion and conclude in Section~\ref{sec:Conclusion}.

%%%%%%%%%%%%%%%%%%%%%%%%%%%%%%%%%%%%%%%%%%%%%%%%%%%
%%%%%%%%%%%%%%%%%%%%%%%%%%%%%%%%%%%%%%%%%%%%%%%%%%%
%%%%%%%%%%%%%%%%%%%%%%%%%%%%%%%%%%%%%%%%%%%%%%%%%%%
\section{Methods}
\label{sec:meth}
%%%%%%%%%%%%%%%%%%%%%%%%%%%%%%%%%%%%%%%%%%%%%%%%%%%
\subsection{Theory}
\label{sec: Theory}
The equations widely used to describe the evolution of planets under the influence of tides are \citep[e.g.][]{Barnes2008, LithwickWu2012,Lee2013}:
\begin{equation}
\dot{e} = -\frac{9}{2}\pi \frac{k}{Q} \frac{1}{m_p} \sqrt{\frac{GM^3}{a^3}} \left(\frac{r_p}{a} \right)^5 e
\label{eq:tidee}
\end{equation}
\begin{equation}
\dot{a} = -9\pi \frac{k}{Q} \frac{1}{m_p} \sqrt{\frac{GM^3}{a^3}} \left(\frac{r_p}{a} \right)^5 e^2 a,
\label{eq:tidea}
\end{equation}
where $a$ is the semi-major axis, $e$ is the eccentricity, $k$ is the planet's Love number, $Q$ is the planet's \textit{tidal quality factor} \citep{Goldreich1966}, $m_p$ and $r_p$ are the planet's mass and radius, respectively, $M$ is the stellar mass, and $G$ is the gravitational constant. 
From Eqs.~\ref{eq:tidee} and~\ref{eq:tidea} it follows that $\dot{e}$ and $\dot{a}$ are related by:
\begin{equation}
\frac{\dot{a}}{a} = 2e\dot{e}
\label{eq:LiWu}
\end{equation}

Which arises from conservation of orbital angular momentum.
We are interested in finding a relationship between the total migration of the system and its state variables (such as $a$ and $e$), which can be obtained by integrating Eq.~\ref{eq:LiWu}: 
\begin{equation*}
\int_{a_f}^{a_i} \frac{da}{a} = 2\int_{e_f}^{e_i} ede
\end{equation*}
\begin{equation}
a_f  = a_i \exp(- e_i^2 + e_f^2),
\label{eq:ae}
\end{equation}
where subscripts $i$ and $f$ refer to the initial and final states, respectively. 

Eq.~\ref{eq:ae} is surprisingly simple -- if we know the initial states $e_i$ and $a_i$ of a planetary body, as well as the final eccentricity $e_f$, we can predict its final position. 
Eq.~\ref{eq:ae} is independent of tidal response parameters ($k$, $Q$), the length of time considered, the mass and radii of the star/planet, etc. 
These additional factors affect the timescale by which the body arrives at its final position, $a_f$, but not the final position of the body itself. 

Let us measure the spacing of a planet pair close to a $j:j+1$ MMR by defining 
\begin{equation}
\Delta \equiv \frac{P_{out}}{P_{in}} - \frac{j+1}{j},
\label{eq:Delta}
\end{equation}
where $P$ is the orbital period, and the subscripts $in$ and $out$ refer to the inner and outer planet, respectively. 
For the 2:1~MMR, $j= 1$. 
This definition of $\Delta$ is the same as \citet{Lee2013}, and twice the value used by \citet{LithwickWu2012}. 

Using the fact that $P \propto a^{3/2}$, and substituting Eq.~\ref{eq:ae} into Eq.~\ref{eq:Delta}, we recast $\Delta$ in terms of $a_i$ and $e_i$ for the near resonant pair (setting $j=1$):
\begin{equation}
\Delta = \left(\frac{a_{out,f}}{a_{in,f}} \right)^{3/2} - 2
\label{eq:DeltaInitial}
\end{equation}
\begin{equation*}
\Delta = \left(\frac{a_{out,i}\exp(-e_{out,i}^2 + e_{out,f}^2)}{a_{in,i}\exp(-e_{in,i}^2 + e_{in,f}^2)} \right)^{3/2} - 2
\end{equation*}
We make the assumption that $e_{out,i} \approx e_{out,f}$~(see Section~\ref{sec:Discussion} for a discussion).
After simplifying, we get:
\begin{equation}
\Delta = \left(\frac{a_{out,i}\exp(e_{in,i}^2 - e_{in,f}^2)}{a_{in,i}} \right)^{3/2} - 2.
\label{eq:DeltaFinal}
\end{equation}
Thus, Eq.~\ref{eq:DeltaFinal} relates the final spacing of the system, $\Delta$, to the initial conditions of the system ($a_i$ and $e_i$) and the final eccentricity of the inner planet, $e_{in,f}$.
Eq.~\ref{eq:DeltaFinal} assumes that the angular momentum of each individual planet is conserved, which in general is not true for multi-planet systems (only total angular momentum is).
Other works \citep{LithwickWu2012,Batygin2013} have accounted for this fact, resulting in more accurate equations for the evolution of a planet pair.
We now aim to compare Eq.~\ref{eq:DeltaFinal} to numerical simulations, and estimate its accuracy for \kep{} systems with $\Delta < 0.06$ of the 2:1 MMR. 
We first outline our experimental setup.

%%%%%%%%%%%%%%%%%%%%%%%%%%%%%%%%%%%%%%%%%%%%%%%%%%%
\subsection{Experimental Setup}
\label{sec:expsetup}
Numerical simulations are carried out using the Wisdom $\& $ Holman integration scheme \citep{Wisdom1991}, implemented via REBOUND \citep{Rein2012}. 
Our sample consists of \kep{} systems near the 2:1 MMR with $\Delta < 0.06$, which we call \textit{near-resonant pairs}.
Our choice of $\Delta < 0.06$ stems from a natural cutoff where the excess of near-MMR ends, as shown in Figure~\ref{fig:MMR}. 
We exclude near-resonant pairs interior to the 2:1 MMR, since tidal forces appear only to increase planet separation with time \citep[e.g.][]{LithwickWu2012}.
In addition, we also exclude near-resonant pairs that are \textit{complex}. 
By complex, we mean dynamically involved in an additional (near) resonance (e.g. Laplace resonance), and/or containing an additional planet orbiting between the near-resonant pair. 
Our exclusion of complex resonant systems decreases the number of \kep{} systems by 6.
Lastly, we also remove the Kepler-11 system, which does not contain a complex resonance, but does go unstable on short ($\sim$Myr) timescales when placed into resonance. 
This leaves us with 27 \kep{} systems, and their properties are displayed in Table~\ref{tab:kepsys}.
We remind the reader that for each \kep{} system we simulate the entire system, not just the near-resonant planets. 

Many \kep{} planets do not currently have measured mass values, so we assign planet masses using Eq. 3 from \citet{Weiss2014} for planets $r_p < 4 r_{\oplus}$:  $(m_p/m_{\oplus}) = 2.69(r_p/r_{\oplus})^{0.93}$, and assume a density of Jupiter for planets $r_p > 4r_{\oplus}$.  
In addition, we also input mass values from the transit-timing variation study of \citet{Hadden2014} where applicable. 
For stars without measured stellar masses, we assume $M/M_{\odot} = (R/R_{\odot})^{1.25}$, derived from \cite{Demircan1991}.
For simplicity, we also assume that the inclination of our \kep{} planets is zero.

We assign k/Q values following a similar prescription as \cite{Lee2013} by assigning the most generous values possible. For Earth-like rocky planets, k/Q($r_p/r_{\oplus} < 2$) = 1/40, for planets smaller than Jupiter, k/Q($2 < r_p/r_{\oplus} < 10$) = 1/22000, and for Jupiter-sized giant gaseous planets, k/Q($r_p/r_{\oplus} > 10$) = 1/54000. 

To speed up simulation time, we increase $k/Q$ by a factor of $A_{k/Q}$ (or alternatively this could also be interpreted as increasing tidal strength). 
This tactic has been used by other scientists \citep[e.g.][]{Delisle2014}, and is valid as long as $\tau_e$ is much longer than the planet's eccentricity libration time.
We simulate our \kep{} sample for 50 Myr, and use $A_{k/Q} = 200$, giving a total simulation time of $T = 10$ Gyr.

To begin our simulations, we place each planet at a distance of 1.15$a_{obs}$, where $a_{obs}$ is the observed semi-major axis value from the \kep{} catalog, and assign $e_i = 0.01$. 
We then migrate each planet (in a type-I fashion) back to its original starting position $a_{obs}$ except for of the outer near-resonant planet, which instead migrates a distance of $a_{obs} + \Delta$, forcing the near-resonant pair into a 2:1 MMR.

Each planet migrates for time $t_{mig}$ at rate $\dot{a} = an\mu^{4/3}/C$, where $n$ is the mean motion of the inner planet, $\mu$ is the planet/star mass ratio, and $C$ is a constant. 
Lower values of $C$ cause the outer planet to encounter the MMR sooner, allowing time for both planets to migrate together in resonance, which increases eccentricity to a desired value \citep{Lee2002}. 
For the restricted 3-body problem \citet{Goldreich2014} guarantee capture into resonance if $C_{out} > 3.75$, $e_{in} < (\mu_{out}/j)^{1/3}$ but we use a more conservative value of $C_{out} = 6$ as well as perform numerical tests to ensure that overstabilities do not occur on Myr timescales. 

Defining $K \equiv \frac{\dot{e_i}/e_i}{\dot{a_i}/a_i}$, we use a default $K = 100$ when migrating planets into resonance but also experiment with $K = 10$. 
$K$ (along with $m_p/M$) affects the resulting equilibrium eccentricity, but does not affect tidal evolution, and thus does not affect our main conclusions. 
At this point, initial eccentricities of our simulated \kep{} systems range from $0.05 < e_i < 0.25$, depending on the value of $C$, $K$, etc.

After time $t_{mig}$, migration is quenched over a timescale of $t_{mig}/3$ by letting $\tau_a \rightarrow \infty$ and $\tau_e \rightarrow \infty$, where $\tau_a \equiv - a/\dot{a}$ and $\tau_e \equiv - e/\dot{e}$.
It is at this point that tides are turned on, and the system evolves under the influence of Eq.~\ref{eq:tidee} and \ref{eq:tidea} for the remainder of the simulation. 

%%%%%%%%%%%%%%%%%%%%%%%%%%%%%%%%%%%%%%%%%%%%%%%%%%%
%%%%%%%%%%%%%%%%%%%%%%%%%%%%%%%%%%%%%%%%%%%%%%%%%%%
%%%%%%%%%%%%%%%%%%%%%%%%%%%%%%%%%%%%%%%%%%%%%%%%%%%
\section{Results}
\label{sec:results}
%%%%%%%%%%%%%%%%%%%%%%%%%%%%%%%%%%%%%%%%%%%%%%%%%%%
\subsection{Theory vs. Numerics}
\label{sec:th_v_num}
\begin{figure}
\centerline{\includegraphics[trim=0cm 0cm 0.5cm 1cm, scale=0.48]{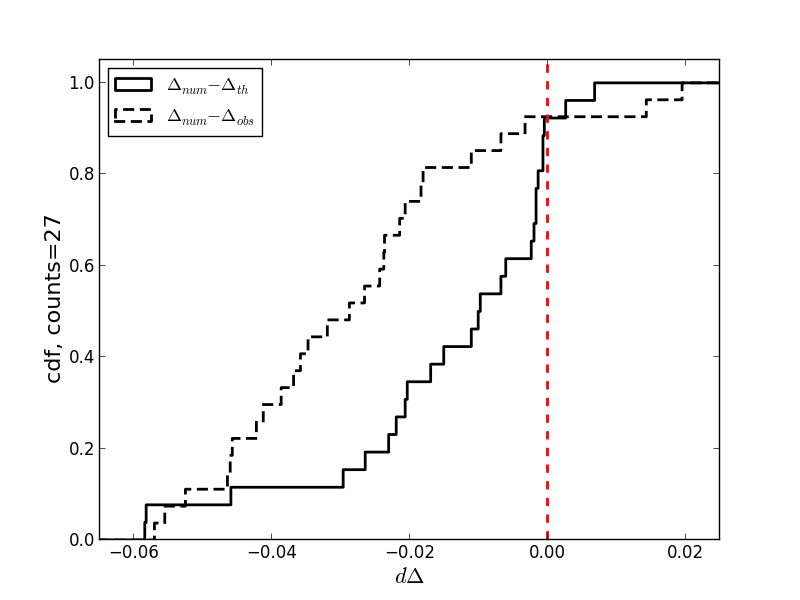}}
\caption{Cumulative distribution function (CDF) of our results. The solid line shows $\Delta_{num} - \Delta_{th}$, the difference between the theoretical and simulated planet separations after $T$ = 10 Gyr. The dashed line shows $\Delta_{num} - \Delta_{obs}$, the difference between our numerical simulations and the observed \kep{} spacing.
}
\label{fig:th_v_num}
\end{figure}

We compare our theoretical predictions of tidal evolution, $\Delta_{th}$ (Eq.~\ref{eq:DeltaFinal}), to our numerical simulations, $\Delta_{num}$ (Eq.~\ref{eq:DeltaInitial}). 
The difference, $\Delta_{num} - \Delta_{th}$, is displayed as a solid line in Figure~\ref{fig:th_v_num}, and expressed as a cumulative distribution (CDF).
For all but 2 systems we see that $\Delta_{num} - \Delta_{th} < 0$, indicating that our theoretical predictions $\Delta_{th}$ consistently over-predict our numerical results, $\Delta_{num}$. 
These 2 exceptions, Kepler-32 and Kepler-221, are expected due to the fact that $T/\tau_e \sim 1000$, which allows extensive $\Delta \propto t^{1/3}$ resonant repulsion growth \citep{LithwickWu2012} in time $T$ which is not accounted for in Eq.~\ref{eq:DeltaFinal}.
Otherwise, we find that Eq.~\ref{eq:DeltaFinal} consistently overestimates the true evolution of the system.

The reason for this $\Delta_{num} - \Delta_{th} < 0$ trend is due to \textit{resonant tugging}, an effect present in the numerics but not captured by our theoretical predictions.
Resonant tugging acts to keep planets closer together than theory would predict. We explore resonant tugging in the next section. 

For the same simulations we also plot $\Delta_{num} - \Delta_{obs}$ as a dashed line in Figure~\ref{fig:th_v_num}, which is the difference between our numerical results, $\Delta_{num}$, and the observed spacing of \kep{} planets seen today, $\Delta_{obs}$ (Eq.~\ref{eq:Delta}). 
As is clearly shown, $\Delta_{num} - \Delta_{obs} < 0$ for all but two systems, suggesting that tides alone cannot explain the observed spacing of \kep{} planets.
These two exceptions are again, Kepler-32 and Kepler-221, and are exceptions for the same reason as above. 
Since $\Delta_{obs} = 0.038$ and $0.035$ for Kepler-32 and Kepler-221, respectively, using a lower $\Delta$ cutoff for our \kep{} sample (e.g. $\Delta_{obs} < 0.03$) would not have changed our result that tides cannot explain near-resonant pairs.

Although highly suggestive, this result does not conclusively disprove tides as the primary evolving mechanism, since higher $e_i$ could cause more migration in time $T$ (see Eq.~\ref{eq:ae}), and the median eccentricity for our simulations is $e_{in,i} = 0.14$. 
Higher values of $e_{in,i}$ are possible. 
Instead of numerically exploring every possible $e_{in,i}$ value, however, we instead reverse the argument in Section~\ref{sec:mine} and calculate the minimum $e_{in,i}$ required to explain $\Delta_{obs}$.
First, however, we explore resonant tugging. 

%%%%%%%%%%%%%%%%%%%%%%%%%%%%%%%%%%%%%%%%%%%%%%%%%%%
\subsection{Resonant Tugging}
\label{sec:restugg}
\textit{Resonant tugging} affects planets in MMR subjected to energy dissipation (e.g. tidal), with moderate to high eccentricity. 
When these conditions are met and the inner planet migrates inwards (trying to leave the resonance) it tugs the outer planet inwards along with it, transferring dissipative forces from the inner to the outer planet.
The result is that the inner planet migrates less than expected, the outer planet migrates more than expected, and the planets are closer together than theory would have predicted.

\begin{figure}
\centerline{\includegraphics[trim=0cm 0cm 1cm 1cm, scale=0.48]{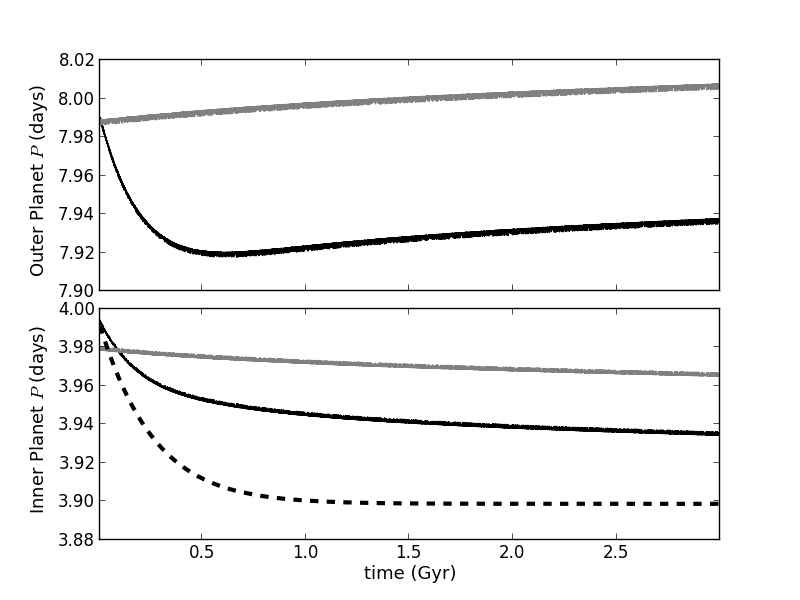}}
\caption{ Two test cases illustrating resonant tugging and repulsion.
The top and bottom panel shows the period evolution of the inner and outer planet, respectively.
For the black curve $e_{in,i} = 0.125$, while for the grey curve $e_{in,i} = 0.018$. 
The dotted black curve shows the numerical trajectory of the inner planet ($e_{in,i} = 0.125$) in the absence of the outer planet.
}
\label{fig:repulse}
\end{figure}

Figure~\ref{fig:repulse} illustrates resonant tugging for a pair of $m = 10^{-4}M$ planets in MMR.  
For the black curve $e_{in,i} = 0.125$, for the grey curve $e_{in,i} = 0.018$, and otherwise the initial conditions of each test case are the same.
In both cases we allow \textit{only the inner planet} to evolve under the influence of tides.
The top and bottom panels show the period evolution of the outer and inner planets, respectively, and the dotted curve in the bottom panel shows evolution of the inner planet in the absence of the outer planet (also $e_{in,i} = 0.125$).

Resonant tugging is exhibited in the first 0.5 Gyr of evolution for the black curve, showing how tidal forces affecting the inner planet also affect the outer planet by dragging it inwards too. 
Comparing the solid and dotted black curves in the bottom panel of Figure~\ref{fig:repulse}, we see that in the presence of the outer planet, the inner planet migrates much less than expected due to resonant tugging. 
Since the outer planet has also migrated inwards more than expected, the result is that $\Delta_{black}$ has grown very little over the first 0.5 Gyr of evolution ($\Delta_{black} = 0.004$ after 0.5 Gyr). 

Within the framework of resonant tugging, the outer planet can be thought of as a massive anchor -- as $m_{out} / m_{in} \rightarrow \infty$ the inner planet has an increasingly difficult time migrating both bodies inwards, leading to a pair of (relatively) stationary planets. 
Conversely, as $m_{out} / m_{in} \rightarrow 0$ the inner planet has an easier time migrating both bodies inwards, and the trajectory of the inner planet will approach its single-planet trajectory (i.e. the dotted black line in Figure~\ref{fig:repulse}).

We now briefly contrast resonant tugging from resonant repulsion \citep[first described by][]{LithwickWu2012}.
Qualitatively, the two main differences between resonant tugging and resonant repulsion are:
\begin{enumerate}
\item Resonant tugging exclusively affects planets in resonance (resonant angle(s) librating) with moderate to high eccentricity, while resonant repulsion affects planets both in and close to resonance, and is most noticeable in the $e \ll 1$ limit. 

\item Resonant tugging decreases $a_{in}$ and decreases $a_{out}$, while resonant repulsion decreases $a_{in}$ and increases $a_{out}$. 
\end{enumerate}

These two differences are illustrated in Figure~\ref{fig:repulse}.
The grey curve, which has low initial eccentricity ($e_{in,i} = 0.018$) exhibits pure resonant repulsion -- a decrease in $a_{in}$ and an increase in $a_{out}$, while the black curve first exhibits a shorter period of resonant tugging followed by resonant repulsion.
The transition from resonant tugging to resonant repulsion for the black curve occurs after 0.5 Gyr, when the eccentricity of the inner planet has dropped to a low ($e_{in} = 0.035$) value.

In Figure~\ref{fig:repulse}, after $\sim1$ Gyr, $\Delta_{black} < \Delta_{grey}$, but $\dot{\Delta}_{grey} = \dot{\Delta}_{black}$, showing how resonant tugging can permanently stunt the growth of $\Delta$.
We also see for the black curve that $a_{out,f} < a_{out,i}$, and it is only after many $\tau_e$ timescales that the outer planet can recover (or exceed) its initial position via resonant repulsion. 
Furthermore, since to first order (when $T/\tau_e \ll 1000$) $\dot{\Delta} \propto \tau_{a,in} \propto e_{in,i}^2$ one would naively expect the black curve to experience $e^2$ more $\Delta$ growth in time $T$, yet we actually find $\Delta_{black} < \Delta_{grey}$, illustrating just how significant resonant tugging can be.
This means that the high eccentricity tidal mechanism suggested by \citet{Delisle2014} does not work for planets in resonance, since (contrary to expectation) high eccentricity actually stunts the growth of $\Delta$, not accelerates it. 

The results shown in Fig.~\ref{fig:th_v_num} are due to resonant tugging, and is supported by the fact that for every simulated \kep{} system we find $a_{out,f} / a_{out,i} < 1$. 
Since resonant repulsion can only increase $a_{out}$ (as shown in Fig.~\ref{fig:repulse}), $a_{out,f} / a_{out,i} < 1$ can only be due to resonant tugging since inward tidal migration is a negligible contribution for the outer planet.
The analytic aspects of resonant tugging will be studied in more depth in future works.

%%%%%%%%%%%%%%%%%%%%%%%%%%%%%%%%%%%%%%%%%%%%%%%%%%%
\subsection{Minimum Eccentricity to Explain $\Delta_{obs}$}
\label{sec:mine}
Since we found in Section~\ref{sec:th_v_num} that $\Delta_{th}$ consistently over-predicts the amount of tidal migration (due to resonant tugging), we can use it as an upper limit predictor of tidal evolution, assuming that near-resonant pairs started in 2:1 MMR and evolved to their present locations. 
Starting from Eq.~\ref{eq:LiWu}, and using the same logic as Section~\ref{sec:th_v_num} we calculate the minimum eccentricity required by the inner planet to achieve the observed spacing seen today, $\Delta_{obs}$, after $T$ years (see Appendix~\ref{app:min_e} for detailed calculations):
\begin{equation}
e_{in,i} = \sqrt{\frac{\ln[(\frac{a_i}{a_f})_{out} (\frac{\Delta_{obs} + 2}{2})^{2/3}]}{1 - \exp(-2T/\tau_{e,in})}}
\label{eq:emin}
\end{equation}
Thus, for a pair of planets starting in 2:1 MMR, if we know the observed spacing today, $\Delta_{obs}$, the number of $\tau_{e,in}$ damping timescales in time $T$, and estimate the amount of migration done by the outer planet in time $T$, $(a_i /a_f)_{out}$, we can calculate the {\it minimum} initial eccentricity that the inner planet must have in order to arrive at the current spacing, $e_{in,i}$. 
Since we assume a starting position of exact 2:1 commensurability, Equation~\ref{eq:emin} is only tailored for $\Delta_{obs} > 0$. Tides cannot decrease planet spacing over time.

\begin{figure}
\centerline{\includegraphics[trim=0cm 0cm 1cm 1cm, scale=0.48]{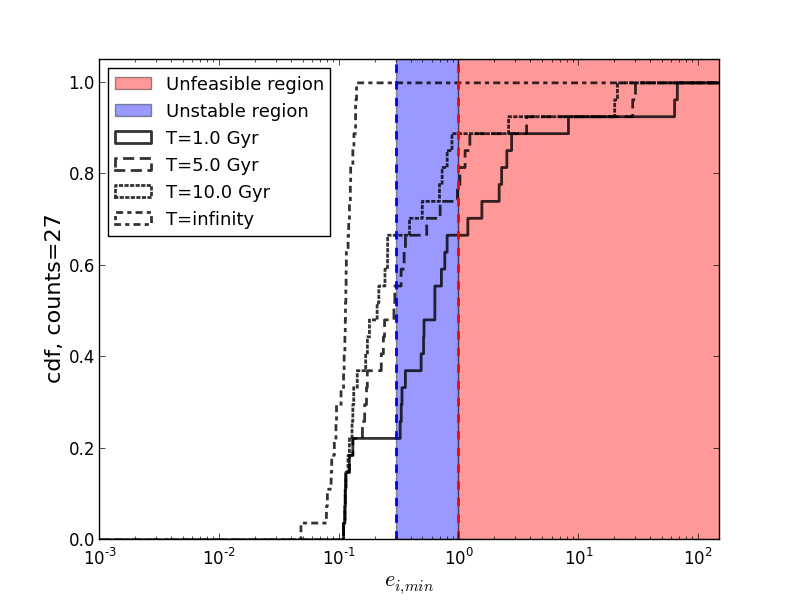}}
\caption{Three CDFs showing the theoretical minimum eccentricity required by the inner planet in order to achieve the observed $\Delta$ spacing seen by \kep{} planets today. The solid, dashed, dotted lines represent $T = 1, 5, 10$ Gyr tracks, respectively, while the dash-dotted line represents $T \rightarrow \infty$. In all calculations we assume the outer planet remains stationary, i.e. $(a_i /a_f)_{out}$ = 1. The red shaded region marks the unphysical region where the eccentricity is larger than unity. The blue region marks the region where most systems undergo a dynamical instability. }
\label{fig:emin}
\end{figure}

Figure~\ref{fig:emin} shows CDFs of Eq.~\ref{eq:emin} applied to our \kep{} sample for $T$ = 1 (solid), 5 (dashed), and 10 (dotted) Gyr with the outer planet remaining stationary, i.e. $(a_i /a_f)_{out} = 1$ (see Section~\ref{sec:Discussion} for a discussion).
In addition we plot a $T \rightarrow \infty$ curve as a dash-dotted line, which the other curves converge to.
The shaded red region marks where $e_{in,i} \geq 1.0$, while the blue marks an unstable region where eccentricities are unlikely to exist.
We construct the blue region by numerically finding the maximum eccentricity allowed before $> 50\%$ of our \kep{} systems go unstable within 2 Myr (see Appendix~\ref{app:emax} for details). 
We find this maximum eccentricity to be $\approx 0.3$ (the left boundary of the blue shaded region in Figure~\ref{fig:emin}).

The three $T < \infty$ curves in Figure~\ref{fig:emin} lie largely in the red and blue shaded regions, indicating that the required eccentricities to explain $\Delta_{obs}$ are unreasonable. 
We thus conclude that most \kep{} systems cannot be explained by tides alone.
Even for the $T=10$ Gyr curve, an optimistic estimate for the age of many \kep{} systems, about $35\%$ of systems still cannot be explained due to tides alone. 
Clearly, another mechanism is needed to explain the near-resonant 2:1 MMR pairs.

%%%%%%%%%%%%%%%%%%%%%%%%%%%%%%%%%%%%%%%%%%%%%%%%%%%
%%%%%%%%%%%%%%%%%%%%%%%%%%%%%%%%%%%%%%%%%%%%%%%%%%%
%%%%%%%%%%%%%%%%%%%%%%%%%%%%%%%%%%%%%%%%%%%%%%%%%%%
\section{Discussion}
\label{sec:Discussion}
A number of assumptions have been made in constructing Figure~\ref{fig:emin} which only strengthen our conclusion that planets close to the 2:1 MMR cannot be explained due to tides alone. 

First and foremost, from Sections~\ref{sec:th_v_num} and~\ref{sec:restugg} we have shown that resonant tugging causes our theoretical predictions of $\Delta$ to be overestimates, and thus Eq.~\ref{eq:emin} underestimates the minimum $e_{in,i}$ required to explain the current observed spacing. 
This discrepancy becomes most pronounced when $m_{out}/m_{in} > 1$, which is the case for many \kep{} systems in our sample.
In particular we showed that resonant tugging can stunt the evolution of planets away from MMR by roughly $e^2$ (see Sec~\ref{sec:restugg}), and this stunted evolution is not accounted for in Figure~\ref{fig:emin}.  

Second, we have assumed optimistic $k/Q$ values which allow for more migration in time $T$. 
Although \textit{some} planets could have such generous values, it is unlikely that all of them do. 
Since $\dot{a} \propto (k/Q)e^2$, as $k/Q$ decreases $e_{in,i}$ must increase in order for the planets to achieve the same observed $\Delta$ in time $T$.

Third, our estimate of $e_{max} = 0.3$ (blue-shaded region in Figure~\ref{fig:emin}) is very likely an overestimate. 
For example, \cite{Pu2015} found that the mean eccentricity of high-multiple Kepler planets must not exceed $e_{mean} = 0.02$ to guarantee long-term dynamical stability. 
Also, from simple orbit crossing arguments when $e > 0.23$ the perihelion of the outer planet crosses the aphelion of the inner planet for a 2:1 MMR, and long-term stability can no longer be guaranteed.
Even if it were possible for \kep{} systems to remain stable with high ($> 0.3$) eccentricities, it is unclear what kind of mechanism could consistently generate them for our \kep{} sample.

There are other assumptions we made throughout this paper which we now justify.
For the results presented in Fig.~\ref{fig:emin} we assumed that the outer planet remains stationary, i.e. $(a_i /a_f)_{out} = 1.00$. 
This is a reasonable assumption since, referring to our $T = 10$ Gyr numerical simulations as a benchmark, the median value of $(a_i /a_f)_{out} = 1.001 \approx 1$.
Increasing the value of $(a_i /a_f)_{out}$ shifts our CDFs in Fig.~\ref{fig:emin} further into the blue/red instability region. 

In Eq.~\ref{eq:DeltaFinal} we assumed that $e_{out,i} \approx e_{out,f}$ which essentially states that the the initial and final positions of the outer planet are the same. 
As stated in the previous paragraph, since we found numerically from our simulations that the median $(a_i /a_f)_{out} \approx 1$, this assumption is reasonable. 

In constructing Fig.~\ref{fig:emin} we have used Eq.~\ref{eq:emin}, which is Eq.~89 from \cite{Delisle2014}, who argued that for moderate to high eccentricities ($e_{in,i} \geq 0.15$), many of the near-resonant pairs could in fact be explained by tides. 
There are a number of differences between our analysis and theirs however. 
First (and most importantly), their estimates are based on theoretical predictions (i.e. resonant tugging unaccounted for), and we have shown in Section~\ref{sec:restugg} that the growth of $\Delta$ is significantly stunted when resonant tugging is accounted for, especially when $m_{out} / m_{in} > 1$.
Second, their analysis assumes that $T \rightarrow \infty$, while we restrict to $T = 1, 5$ and $10$ Gyr. 
Looking at Fig~\ref{fig:emin}, we see that the $T \rightarrow \infty$ curve tells a very different story than the $T = 1, 5$ and $10$ Gyr curves, and the conclusion of whether or not tides can explain $\Delta_{obs}$ is certainly time dependent.
Lastly, \citeauthor{Delisle2014} assumes $\Delta = 0.03$ for all systems, while we use the system specific $\Delta$ values.

As a consistency check for our results, we perform the same set of experiments (including our tests of resonant tugging) using a different version of tides, implementing them in terms of forces (as opposed to orbital elements like in Eq.~\ref{eq:tidee} and~\ref{eq:tidea}) according to \cite{Papaloizou2000}:
\begin{equation}
\vec{a}_{damp} = -2\frac{(\vec{v} \cdot \vec{r}) \vec{r}}{r^2 \tau_e}
\label{eq:tideF}
\end{equation}
where $\vec{a}_{damp}$ is the damping acceleration, $\vec{v}$ is the velocity, $\vec{r}$ is the position, $r$ is the scalar position and $\tau_e \equiv - e/\dot{e}$ as before. 
Whenever a planet receives a ``kick'' in the WH integration scheme, an additional kick of $a_{damp}$ is supplied to account for tides. 
We find our overall conclusions unaffected using this implementation of tides.

We have omitted complex resonances from our analysis because their behaviour is much more unpredictable. 
For simple resonances we found that $\Delta_{th} > \Delta_{num}$, but some complex resonances violated this relationship. 
The reasons are currently unknown, but will be more thoroughly investigated in future works. 

Lastly, it should be mentioned that spin tides were omitted from this analysis, which arise when the spin rate of the host star $\Omega_*$ is different than the mean motion $n$ of the orbiting planet (responsible for the Moon's recession from Earth over time). 
Spin rates of \kep{} stars are largely unknown, as well as the evolution of these spin rates, $d\Omega_*/dt$. 
Depending on the sign of $(\Omega_* - n)$, spin tides can induce inward or outward migration. 
It is thus a non-trivial process to determine what the affect of spin tides might be on the evolution of a system. 
The equation governing spin tide migration is \citep{MurrayDermott}:
\begin{equation}
\dot{a}_p = {\rm sign}(\Omega_* - n_p)\frac{3k_*}{Q_*}\frac{m_p}{M}\left(\frac{R}{a_p}\right)^5 n_pa_p
\label{eq:spina}
\end{equation}
where the subscripts $p$ and $*$ refer to the planet and star, respectively. 
We can estimate the relative strength of eccentricity tides (Eq.~\ref{eq:tidea}) to spin tides (Eq.~\ref{eq:spina}):
\begin{equation*}
\frac{\dot{a}_{ecc}}{\dot{a}_{spin}} = \frac{9\pi\frac{k_p}{Q_p} \sqrt{\frac{GM^3}{a_p^3}} \frac{e_p^2a_p}{m_p}}{ 3\frac{k_*}{Q_*}\frac{m_p}{M}\left(\frac{R}{a_p}\right)^5 n_pa_p} = 3\pi\frac{k_p}{k_*}\frac{Q_*}{Q_p}\left(\frac{M}{m_p} \right)^2 \left(\frac{r_p}{R} \right)^5 e^2
\end{equation*}
Interestingly enough, the relative strength of spin vs. eccentricity tides is independent of the semi-major axis.
Assigning typical values from \citet{WuMurray2003} for $\Omega_*$, $(k/Q)_*$, $M$ and $R$, and assuming a typical $\sim 4m_{\oplus}$ planet we get:
\begin{equation}
\frac{\dot{a}_{ecc}}{\dot{a}_{spin}} \sim 30
\end{equation}
Combining this with the fact that $\Omega_*$ and $d\Omega_*/dt$ are largely unknown for \kep{} stars, we felt justified omitting spin tides from our analysis.

\section{Conclusion}
\label{sec:Conclusion}
In conclusion, we have investigated 27 \kep{} systems containing 2:1 near-resonant pairs, and find that tides alone cannot explain their current observed spacing, $\Delta_{obs}$. 
In Figure~\ref{fig:emin} we calculated the minimum theoretical eccentricity required by the inner planet to explain $\Delta_{obs}$ and found that for a large number of systems $e_{in,i} > 0.3$, which from simple dynamical arguments is not a reasonable eccentricity for \kep{} planets to have.
Furthermore, our numerical study of resonant tugging reveals that our theoretical predictions of $e_{in,i}$ are optimistic estimates, and in cases where $m_{out}/m_{in} > 1$, significantly so.
A number of other assumptions made throughout the paper contribute to these optimistic estimates.

As a numerical compliment to our theoretical investigation, we simulated our \kep{} sample for 10 Gyr with a median eccentricity of $e_{in,i} = 0.14$, and found only two systems, Kepler-32 and Kepler-221, that migrated to $\Delta_{obs}$ (dotted line in Figure~\ref{fig:th_v_num}).
Clearly, another mechanism is required to explain the excess of \kep{} systems exterior to the 2:1 MMR. 

%%%%%%%%%%%%%%%%%%%%%%%%%%%%%%%%%%%%%%%%%%%%%%%%%%%%%%%%%%%%%

\section*{Acknowledgments}
This research has made use of the NASA Exoplanet Archive, which is operated by the California Institute of Technology, under contract with the National Aeronautics and Space Administration under the Exoplanet Exploration Program. This research is supported by an NSERC CGS D award to AS and an NSERC Discovery Grant RGPIN-2014-04553. We thank the reviewer for their very insightful comments and thorough review.

%%%%%%%%%%%%%%%%%%%%%%%%%%%%%%%%%%%%%%%%%%%%%%%%%%%%%%%%%%%%%
\bibliographystyle{apj}

\bibliography{paperref.bib}

\appendix
%%%%%%%%%%%%%%%%%%%%%%%%%%%%%%%%%%%%%%%%%%%%%%%%%%%
%%%%%%%%%%%%%%%%%%%%%%%%%%%%%%%%%%%%%%%%%%%%%%%%%%%
%%%%%%%%%%%%%%%%%%%%%%%%%%%%%%%%%%%%%%%%%%%%%%%%%%%
\section{Calculation of Minimum Eccentricity Required for Inner Planet Given $\Delta_{obs}$ and $T$ Years}
\label{app:min_e}

Starting from Eq.~\ref{eq:LiWu}, we now derive Eq.~\ref{eq:emin}:
\begin{equation*}
\frac{\dot{a}}{a} = 2e\dot{e}
\end{equation*}
\begin{equation}
\frac{da}{a} = 2e^2\frac{\dot{e}}{e}dt
\label{eq:daa}
\end{equation}
Since $\tau_e \equiv - \dot{e}/e$ is a constant, the only quantity with a time dependence on the right hand side is $e$.  
Integrating Eq.~\ref{eq:tidee} gives us an expression for $e(t)$:
\begin{equation*}
\frac{de}{e} = -\frac{9}{2}\pi \frac{k}{Q} \frac{1}{m_p} \sqrt{\frac{GM^3}{a^3}} \left(\frac{r_p}{a} \right)^5 dt =  \frac{\dot{e}}{e} dt  = -\frac{1}{\tau_e} dt
\end{equation*}
Since (again) we assume $\tau_e \equiv - e/\dot{e}$ is a constant, the integration is straightforward to yield:
\begin{equation}
e(t) = e_i \exp\left(-\frac{1}{\tau_e} t \right)
\label{eq:et}
\end{equation}

Now that we have eccentricity as a function of time, we can plug Eq.~\ref{eq:et} into Eq.~\ref{eq:daa} and integrate to get:
\begin{equation*}
\int_{a_i}^{a_f} \frac{da}{a} = 2e_i^2 \int_{0}^{T} \frac{-1}{\tau_e} \exp\left(-2\frac{1}{\tau_e} t \right) dt
\end{equation*}
\begin{equation}
\ln(a_i/a_f) = e_i^2 \left(1 - \exp(-2\frac{1}{\tau_e}T) \right)
\label{eq:X}
\end{equation}
We rearrange Eq.~\ref{eq:X} for $e_i$ to get:
\begin{equation}
e_i = \sqrt{\frac{\ln(a_i/a_f)}{1 - \exp(-2T/\tau_e)}}
\label{eq:ei}
\end{equation}
Where all quantities refer to a single planet (e.g. the inner planet). 
To connect Eq.~\ref{eq:ei} to a 2:1 MMR pair, we plug in Eq.~\ref{eq:DeltaInitial} along with the fact that $a_{in,i} = a_{out,i}/2^{2/3}$ (for 2:1 MMR) to get:
\begin{equation*}
e_{in,i} = \sqrt{\frac{\ln(a_i/a_f)_{in}}{1 - \exp(-2T/\tau_{e,in})}}
\end{equation*}
\begin{equation*}
e_{in,i} = \sqrt{\frac{\ln[(\frac{a_{out,i}}{2^{2/3}}) (\frac{(\Delta_{obs} + 2)^{2/3}}{a_{out,f}}) ]}{1 - \exp(-2T/\tau_{e,in})}}
\end{equation*}
\begin{equation}
e_{in,i} = \sqrt{\frac{\ln[(\frac{a_i}{a_f})_{out} (\frac{\Delta_{obs} + 2}{2})^{2/3}]}{1 - \exp(-2T/\tau_{e,in})}}
\end{equation}
Which is the result displayed in Eq.~\ref{eq:emin}.

%%%%%%%%%%%%%%%%%%%%%%%%%%%%%%%%%%%%%%%%%%%%%%%%%%%
%%%%%%%%%%%%%%%%%%%%%%%%%%%%%%%%%%%%%%%%%%%%%%%%%%%
%%%%%%%%%%%%%%%%%%%%%%%%%%%%%%%%%%%%%%%%%%%%%%%%%%%
\section{Maximum Eccentricity From Dynamical Simulations}
\label{app:emax}
\begin{figure}
\centerline{\includegraphics[trim=0cm 0cm 0.5cm 0.5cm, scale=0.48]{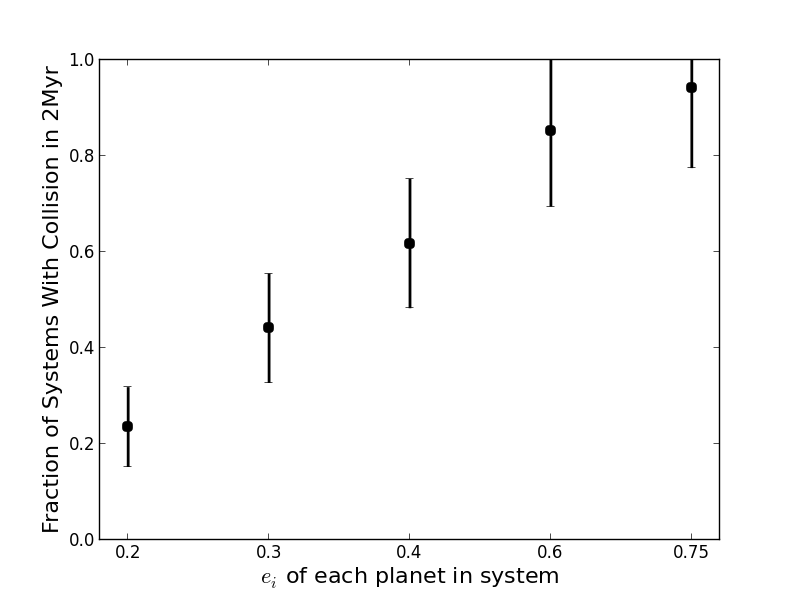}}
\caption{The fraction of \kep{} systems in our sample that go unstable within 2Myr if each planet is given an initial eccentricity of $e_i$. The error bars derived from Poisson statistics.}
\label{fig:emax_dyn}
\end{figure}

In constructing the blue shaded region in Figure~\ref{fig:emin}, we have performed numerical simulations of stability for each Kepler system.
We simulate each Kepler system for 2 Myr, and assign the same initial eccentricity $e_i$ to each planet in the system. 
We do not migrate these planets into resonance, but since (in general) resonance can add a destabilizing effect \citep{Chambers1996, Funk2010, Pu2015}, these simulations can be used as conservative upper limits for the maximum eccentricity that a system can have and remain stable.  
The results are presented in Figure~\ref{fig:emax_dyn}, and error bars are derived from Poisson statistics. 

Each point in Figure~\ref{fig:emax_dyn} represents the fraction of systems in our sample having a collision within 2 Myr.
We can see that for $e_i = 0.3$, about half of the Kepler systems have had a collision. 
Thus from stability arguments, the maximum initial eccentricity that planets in our Kepler sample can have in order for $\sim 50\%$ of them to survive at least 2 Myr is $e_{max} \approx 0.3$.  

%%%%%%%%%%%%%%%%%%%%%%%%%%%%%%%%%%%%%%%%%%%%%%%%%%%
%%%%%%%%%%%%%%%%%%%%%%%%%%%%%%%%%%%%%%%%%%%%%%%%%%%
%%%%%%%%%%%%%%%%%%%%%%%%%%%%%%%%%%%%%%%%%%%%%%%%%%%
\onecolumn
\newpage
\section{Planet Sample}

\begin{center}
%\begin{longtable}{|c|c|c|c|p{2cm}|c|c|c|c|}
\begin{longtable}{rcrrrrrrr}
\caption[Portrait, single page table]{Kepler Systems Used In This Analysis} \\
 
\hline \hline \\[-0.1ex]
   \multicolumn{1}{c}{ {System}} &
   \multicolumn{1}{c}{ {Planet}} &
   \multicolumn{1}{c}{ {$P$ (days)}} &
   \multicolumn{1}{c}{ {near MMR}} &
   \multicolumn{1}{c}{ {$m_p/m_{\oplus}$}} &
   \multicolumn{1}{c}{ {$r_p/r_{\oplus}$}} &
   \multicolumn{1}{c}{ {$N_p$}}  &
   \multicolumn{1}{c}{ {$M/M_{\odot}$}} &
   \multicolumn{1}{c}{ {$R/R_{\odot}$}}  \\[0.5ex]\hline \hline \\[-2ex]
\endfirsthead

%\hline \kep{} System &  Planet Letter &  $P$ (days) & near MMR & $m_p/m_{\oplus}$ &  $r_p/r_{\oplus}$ & $N_p$ & $M/M_{\odot}$ &  $R/R_{\odot}$   \\\hline
KOI-142 & b & 10.95 & \checkmark & $8.7 \pm 2.5$ & $3.82 \pm 0.44$ & 2 & $0.96 \pm 0.04$ & $0.88 \pm 0.03$ \\  
 & c & 22.34 & \checkmark & $198.8 \pm 9.2$ & $6.82 \pm 1.09$ & & & \\  
\hline 
Kepler-120 & b & 6.31 & \checkmark & $  $ & $2.18 \pm 0.22$ & 2 & $  $ & $0.53 \pm 0.03$ \\  
 & c & 12.79 & \checkmark & $  $ & $1.53 \pm 0.11$ & & & \\  
\hline 
Kepler-127 & b & 14.44 & \checkmark & $  $ & $1.42 \pm 0.11$ & 3 & $  $ & $1.36 \pm 0.04$ \\  
 & c & 29.39 & \checkmark & $  $ & $2.62 \pm 0.11$ & & & \\  
 & d & 48.63 & & $  $ & $2.62 \pm 0.11$ & & & \\  
\hline 
Kepler-176 & b & 5.43 & & $  $ & $1.42 \pm 0.76$ & 3 & $  $ & $0.89 \pm 0.46$ \\  
 & c & 12.76 & \checkmark & $  $ & $2.62 \pm 1.31$ & & & \\  
 & d & 25.75 & \checkmark & $  $ & $2.51 \pm 1.31$ & & & \\  
\hline 
Kepler-183 & b & 5.69 & \checkmark & $  $ & $2.07 \pm 0.87$ & 2 & $  $ & $0.96 \pm 0.41$ \\  
 & c & 11.64 & \checkmark & $  $ & $2.29 \pm 0.98$ & & & \\  
\hline 
Kepler-221 & b & 2.8 & \checkmark & $  $ & $1.75 \pm 0.22$ & 4 & $0.72 \pm 0.05$ & $0.82 \pm 0.07$ \\  
 & c & 5.69 & \checkmark & $  $ & $2.95 \pm 0.33$ & & & \\  
 & d & 10.04 & & $  $ & $2.73 \pm 0.22$ & & & \\  
 & e & 18.37 & & $  $ & $2.62 \pm 0.22$ & & & \\  
\hline 
Kepler-244 & b & 4.31 & & $  $ & $2.73 \pm 1.2$ & 3 & $  $ & $0.8 \pm 0.34$ \\  
 & c & 9.77 & \checkmark & $  $ & $2.07 \pm 0.87$ & & & \\  
 & d & 20.05 & \checkmark & $  $ & $2.29 \pm 0.98$ & & & \\  
\hline 
Kepler-25 & b & 6.24 & \checkmark & $9.0 \pm 2.4$ & $2.62 \pm 0.0$ & 3 & $1.19 \pm 0.06$ & $1.31 \pm 0.02$ \\  
 & c & 12.72 & \checkmark & $14.3 \pm 2.7$ & $4.48 \pm 0.0$ & & & \\  
 & d & 123.0 && $89.9 \pm 13.7$ & $5.46 \pm 0.0$ & & & \\  
\hline 
Kepler-267 & b & 3.35 & \checkmark & $  $ & $1.97 \pm 0.11$ & 3 & $0.56 \pm 0.05$ & $0.56 \pm 0.02$ \\  
 & c & 6.88 & \checkmark & $  $ & $2.07 \pm 0.11$ & & & \\  
 & d & 28.46 & & $  $ & $2.29 \pm 0.11$ & & & \\  
\hline 
Kepler-27 & b & 15.33 & \checkmark & $41.8 \pm 5.0$ & $4.04 \pm 0.0$ & 2 & $0.65 \pm 0.16$ & $0.59 \pm 0.15$ \\  
 & c & 31.33 & \checkmark & $21.2 \pm 3.2$ & $4.91 \pm 0.0$ & & & \\  
\hline 
Kepler-272 & b & 2.97 & \checkmark & $  $ & $1.42 \pm 0.76$ & 3 & $0.79 \pm 0.05$ & $0.93 \pm 0.5$ \\  
 & c & 6.06 & \checkmark & $  $ & $1.75 \pm 0.98$ & & & \\  
 & d & 10.94 & & $  $ & $2.29 \pm 1.2$ & & & \\  
\hline 
Kepler-30 & b & 29.33 & \checkmark & $11.3 \pm 1.4$ & $3.93 \pm 0.22$ & 3 & $0.99 \pm 0.08$ & $0.95 \pm 0.12$ \\  
 & c & 60.32 & \checkmark & $640.0 \pm 50.0$ & $12.34 \pm 0.44$ & & & \\  
 & d & 143.34 & & $23.1 \pm 2.7$ & $8.84 \pm 0.55$ & & & \\  
\hline 
Kepler-305 & b & 5.49 & & $10.5 \pm 2.6$ & $3.6 \pm 0.87$ & 3 & $0.76 \pm 0.13$ & $0.79 \pm 0.05$ \\  
 & c & 8.29 & \checkmark & $6.0 \pm 2.4$ & $3.28 \pm 0.76$ & & & \\  
 & d & 16.74 & \checkmark & $  $ & $2.73 \pm 0.44$ & & & \\  
\hline 
Kepler-32 & f & 0.74 & & $  $ & $0.76 \pm 0.11$ & 5 & $0.54 \pm 0.02$ & $0.53 \pm 0.02$ \\  
 & e & 2.9 & \checkmark & $  $ & $1.53 \pm 0.11$ & & & \\  
 & b & 5.9 & \checkmark & $9.4 \pm 3.6$ & $2.18 \pm 0.22$ & & & \\  
 & c & 8.75 & & $7.7 \pm 5.0$ & $1.97 \pm 0.22$ & & & \\  
 & d & 22.78 & & $  $ & $2.73 \pm 0.11$ & & & \\  
\hline 
Kepler-326 & b & 2.25 & \checkmark & $  $ & $1.53 \pm 0.22$ & 3 & $0.98 \pm 0.05$ & $0.8 \pm 0.05$ \\  
 & c & 4.58 & \checkmark & $  $ & $1.42 \pm 0.11$ & & & \\  
 & d & 6.77 & & $  $ & $1.2 \pm 0.11$ & & & \\  
\hline 
Kepler-327 & b & 2.55 & \checkmark & $  $ & $1.09 \pm 0.11$ & 3 & $0.55 \pm 0.05$ & $0.49 \pm 0.02$ \\  
 & c & 5.21 & \checkmark & $  $ & $0.98 \pm 0.11$ & & & \\  
 & d & 13.97 & & $  $ & $1.75 \pm 0.11$ & & & \\  
\hline 
Kepler-328 & b & 34.92 & \checkmark & $28.5 \pm 12.9$ & $2.29 \pm 0.98$ & 2 & $1.15 \pm 0.22$ & $1.06 \pm 0.44$ \\  
 & c & 71.31 & \checkmark & $39.4 \pm 13.6$ & $5.46 \pm 2.29$ & & & \\  
\hline 
Kepler-384 & b & 22.6 & \checkmark & $  $ & $1.09 \pm 0.33$ & 2 & $0.76 \pm 0.05$ & $0.88 \pm 0.25$ \\  
 & c & 45.35 & \checkmark & $  $ & $1.09 \pm 0.33$ & & & \\  
\hline 
Kepler-386 & b & 12.31 & \checkmark & $  $ & $1.42 \pm 0.76$ & 2 & $0.74 \pm 0.05$ & $0.77 \pm 0.43$ \\  
 & c & 25.19 & \checkmark & $  $ & $1.64 \pm 0.87$ & & & \\  
\hline 
Kepler-396 & b & 42.99 & \checkmark & $75.5 \pm 11.8$ & $3.49 \pm 1.31$ & 2 & $0.85 \pm 0.13$ & $1.06 \pm 0.39$ \\  
 & c & 88.5 & \checkmark & $17.9 \pm 2.8$ & $5.35 \pm 1.97$ & & & \\  
\hline 
Kepler-48 & b & 4.78 & \checkmark & $14.3 \pm 4.3$ & $2.18 \pm 0.0$ & 3 & $0.88 \pm 0.06$ & $0.89 \pm 0.05$ \\  
 & c & 9.67 & \checkmark & $9.8 \pm 3.3$ & $3.17 \pm 0.0$ & & & \\  
 & d & 42.9 & & $7.93 \pm 4.6$ & $2.07 \pm 0.11$ & & & \\  
\hline 
Kepler-56 & b & 10.5 & \checkmark & $22.1 \pm 3.9$ & $6.55 \pm 0.33$ & 2 & $1.32 \pm 0.13$ & $4.23 \pm 0.15$ \\  
 & c & 21.4 & \checkmark & $181.0 \pm 21.0$ & $9.83 \pm 0.44$ & & & \\  
\hline 
Kepler-57 & b & 5.73 & \checkmark & $118.1 \pm 24.1$ & $2.18 \pm 0.0$ & 2 & $0.83 \pm 0.05$ & $0.73 \pm 0.0$ \\  
 & c & 11.61 & \checkmark & $7.4 \pm 9.4$ & $1.53 \pm 0.0$ & & & \\  
\hline 
Kepler-79 & b & 13.48 & \checkmark & $  $ & $2.62 \pm 0.76$ & 4 & $1.1 \pm 1.63$ & $1.4 \pm 0.25$ \\  
 & c & 27.4 & \checkmark & $  $ & $2.73 \pm 0.87$ & & & \\  
 & d & 52.09 & & $  $ & $7.64 \pm 1.42$ & & & \\  
 & e & 81.07 & & $  $ & $3.38 \pm 0.66$ & & & \\  
\hline 
Kepler-81 & b & 5.96 & \checkmark & $  $ & $2.4 \pm 0.44$ & 3 & $0.64 \pm 0.38$ & $0.59 \pm 0.03$ \\  
 & c & 12.04 & \checkmark & $  $ & $2.4 \pm 0.33$ & & & \\  
 & d & 20.84 & & $  $ & $1.2 \pm 0.33$ & & & \\  
\hline 
Kepler-83 & d & 5.17 & & $  $ & $1.97 \pm 0.11$ & 3 & $0.66 \pm 0.41$ & $0.59 \pm 0.03$ \\  
 & b & 9.77 & \checkmark & $  $ & $2.84 \pm 0.44$ & & & \\  
 & c & 20.09 & \checkmark & $  $ & $2.4 \pm 0.33$ & & & \\  
\hline 
Kepler-9 & d & 1.59 & & $  $ & $1.64 \pm 0.22$ & 3 & $1.07 \pm 0.05$ & $1.02 \pm 0.05$ \\  
 & b & 19.24 & \checkmark & $80.09 \pm 4.13$ & $9.5 \pm 0.76$ & & & \\  
 & c & 38.91 & \checkmark & $54.35 \pm 4.13$ & $9.28 \pm 0.76$ & & & \\ 
\hline 

\label{tab:kepsys}
\end{longtable}
\end{center}
\twocolumn

\end{document}